\documentclass[preprint,preprintnumbers,amsmath,amssymb]{revtex4}
\usepackage{graphicx}

\begin{document}

\title{Shuttle-mediated proton pumping  \\
across the inner mitochondrial membrane }

\author{ Anatoly Yu. Smirnov$^{1}$, Sergey E. Savel'ev$^{1,2}$, and Franco Nori$^{1,3}$ }

\affiliation{$^1$ Advanced Science Institute, The Institute of Physical and
Chemical Research (RIKEN), \\
Wako-shi, Saitama, 351-0198, Japan \\
$^2$ Department of Physics, Loughborough University, Loughborough
LE11 3TU, UK \\
$^3$ Center for Theoretical Physics, Physics Department, The
University of Michigan, Ann Arbor, MI 48109-1040, USA}

\date{\today}

\begin{abstract}
{Shuttle-assisted charge transfer is pivotal for the efficient energy transduction from the food-stuff electrons to protons in the
respiratory chain of animal cells and bacteria. The respiratory chain consists of four metalloprotein Complexes (I-IV) embedded in
the inner membrane of a mitochondrion. Three of these complexes pump protons across the membrane, fuelled by the energy of
food-stuff electrons. Despite extensive biochemical and biophysical studies, the physical mechanism of this proton pumping is
still not well understood. Here we present a nanoelectromechanical model of the electron-driven proton pump related to the second
loop of the respiratory chain, where a lipid-soluble ubiquinone molecule shuttles between the Complex I and Complex III, carrying
two electrons and two protons. We show that the energy of electrons can be converted to the transmembrane proton potential
gradient via the electrostatic interaction between electrons and protons on the shuttle. We find that the system can operate
either as a proton pump, or, in the reverse regime, as an electron pump. For membranes with various viscosities, we demonstrate
that the uphill proton current peaks near the body temperature $T \approx 37 ^{\circ}$C. }
\end{abstract}


\maketitle

Each mitochondrion has two functionally different membranes: the outer membrane, separating the contents of the mitochondrion from
the cell cytoplasm, and the inner lipid membrane, embedded concentrically inside the outer envelope (see top right of Fig.~1). As
Fig.~1 schematically shows, the inner membrane (in white), which is impermeable to charge and polar particles, divides the
mitochondrial interior in two parts: (i) the central, negatively charged, compartment, referred to as the matrix (N, in yellow),
and (ii) the intermembrane space (P, in blue), which is characterized by a positive charge. The electrical charge difference of
the compartments results in an electrochemical proton gradient $\sim~-$200 mV \cite{Alberts02,Lodish07}, which is used to
synthesize  adenosine triphosphate (ATP) molecules, the main energy currency of the cell. This electrochemical potential is
maintained by a set of proton pumps incorporated into the so-called Complexes I, III, and IV of the respiratory chain
\cite{Mitch61,Mitch76,Schultz01,Hos06,WV07}. Proton pumps employ the energy of electrons, extracted from sugars and fatty acids
after the digestion of food, for the uphill translocation of protons, from the negative (N) to the positive (P) side of the
membrane. In the case of the respiratory chain in mitochondria, the N-side corresponds to the matrix, and the P-side is related to
the intermembrane space (see Fig.~1).

According to chemiosmotic theory \cite{Mitch61,Mitch76} the pumping process in Complexes I and III is organized in \textit{three
functional `loops'}. Each of these loops takes away some of the redox (electron) energy and converts it into energy of protons
\cite{Dutton98}. In the \textit{first loop}, located in Complex I, two electrons are transferred from NADH (a reduced form of
Nicotinamide Adenine Dinucleotide, NAD$^+$) to ubiquinone, pumping four protons in the process.
\cite{Wik84,Brandt06,Verkh08,Friedrich98,Ohnishi05}.

Our primary focus in this work is the \textit{second loop} of the respiratory chain (see Fig.~1). There, a mobile molecule of
ubiquinone, Q$_{10}$ = C$_{59}$H$_{90}$O$_4$, hereafter denoted by Q, takes two electrons from the NADH dehydrogenase enzyme
(Complex I) and two protons from the N-side of the inner mitochondrial membrane, turning Q into a molecule of ubiquinol QH$_2$.
This molecular shuttle diffuses to the P-side of the inner membrane, where it donates both electrons to the cytochrome bc$_1$
(Complex III). The oxidation of ubiquinol (and turning it back to ubiquinone) is accompanied by the pumping of two protons into
the intermembrane space (P-side) against the proton electrochemical potential. Complex III transfers two electrons, acquired from
QH$_2$, to two molecules of cytochrome~$c$, pumping in addition two more protons by means of the Q cycle mechanism (the
\textit{third loop}) \cite{Mitch76,Trump81,Trump90,Brandt96,Hunte03,Armen05,Osyczka04,Crofts06}.

In this work we quantitatively analyze the proton pumping process in the second loop of the respiratory chain. Our goal is to show
that the basic physical effect, related  to the energetically uphill translocation of two protons per two electrons, transferred
from NADH dehydrogenase, can be obtained within a simple nanoelectromechanical model, taking into account the electron-proton
electrostatic coupling on the ubiquinone molecule (shuttle), diffusing inside of the inner membrane between Complexes I and III.
This diffusion is governed by the Langevin equation. In order to describe the process of loading/unloading the shuttle with
electrons and protons, we employ a system of master equations, with position-dependent transition rates between the shuttle and
electron/proton reservoirs. With these equations we analyze the time dependence of the pumping process, as well as the dependence
of particle currents on both temperature and on electron/proton electrochemical gradients. There is a pool of ubiquinone/ubiquinol
molecules in the inner mitochondrial membrane \cite{Dutton98,Schultz01}, but we now only consider the contribution of a single
ubiquinone molecule to the electron and proton translocation process. Because of this, the actual values of the electron and
proton fluxes should be higher than the values calculated below. Note also that a similar proton pumping mechanism can work in
photosynthetic systems \cite{Alberts02, Lodish07,Armen05}.

\subsection{Model}

The mitochondrion can be considered as a nonequilibrium open system \cite{Kim07} which has to be continuously supplied with
electrons and protons. Complex I, characterized by a higher electron electrochemical potential, $ \mu_{\rm S} = V_e/2$,  serves as
a \textit{Source} of electrons (S), whereas Complex III, with an electrochemical potential $\mu_{\rm D} = - V_e/2$, acts as a
\textit{Drain} (D), accepting electrons during the process of ubiquinol (QH$_2$) oxidation \cite{Dutton98}. Note that we include
the sign of the electron charge in the definition of the electrochemical potential (see also \textit{Methods}). This means that a
site with a higher electron energy is characterized by a more negative redox midpoint potential $E_{m7}.$ The positive and
negative sides of the membrane are characterized by the proton electrochemical potentials $\mu_{\rm P} = V_p/2$ and $\mu_{\rm N} =
- V_p/2.$ At physiological conditions, both electron ($V_e$) and proton ($V_p$) electrochemical gradients are positive: $V_e > 0,
\; V_p > 0.$

 In our approach, the process of loading (from the source) and unloading (to the drain) the shuttle with electrons
 and with protons (from the N-side to the P-side of the inner membrane) is governed by a set of master equations (see
\textit{Methods}) derived using methods of statistical mechanics and condensed matter theory \cite{Wingr,NEMS,RobPRB08,ASPRE08}.
The shuttle (ubiquinone molecule) has two electron-binding sites, $1_e$ and $2_e$, with energies $\epsilon_{1e}$ and
$\epsilon_{2e}$, as well as two proton-binding sites, $1_p$ and $2_p$, having energies $\epsilon_{1p}$ and $\epsilon_{2p}.$ To
describe this situation quantum-mechanically we introduce 16 occupational states of the shuttle (see \textit{Methods}).Our model
for the nanomechanical proton pump can also work with just one electron and just one proton sites on the shuttle, in contrast to
the kinetic models of pumps in cytochrome $c$ oxidase \cite{Kim07,ASPRE08}. In its completely reduced form of ubiquinol QH$_2$,
the shuttle has a maximal load of two electrons and two protons, whereas in its oxidized quinone form (denoted by Q in Fig.~1) the
shuttle is empty.

 The Coulomb interaction between charges located on the shuttle is given by
the following parameters: the repulsion energy $u_e$ of two electrons, residing on sites $1_e$ and $2_e$, and the repulsion energy
$u_p$ of two protons located on sites $1_p$ and $2_p$. The energy of the Coulomb attraction between an electron occupying the site
$\sigma$ and a proton sitting on the site $\sigma'$ $(\sigma, \sigma'=1,2)$ is given by $u_{\sigma \sigma'}$. In view of the
spatial symmetry of the ubiquinone molecule \cite{Lodish07}, and for the sake of simplicity and without loss of generality, we
describe all electrostatic interactions between electrons and protons on the shuttle by a single electrostatic energy $U_0$:
$u_{\sigma \sigma'}\,=\,u_e\,=\,u_p\,=\,U_0 $, for any combinations of $\sigma$ and $\sigma'$.

We consider here the spatial projection of the shuttle motion on the straight line connecting the Q-binding sites of Complex I and
Complex III. Along this line, the position of the shuttle is characterized by a coordinate $x$. The Brownian motion
\cite{AnnPhys05} of the ubiquinone/ubiquinol molecule along this line is governed by an overdamped Langevin equation (see Eq.~(3)
in \textit{Methods}) with drag coefficient $\zeta$ and diffusion constant $D~=~T/\zeta~(k_B~=~1).$

We find that for the optimal performance of the pump, all eigenenergies of the electron ($ E_e = \epsilon_{\sigma e}$) and proton
($ E_p = \epsilon_{\sigma p}$) sites  on the shuttle ($\sigma = 1,2$) are approximately equal to one half of the charging energy
$U_0: \ E_e = E_p = U_0/2 > 0$.  This energy scale should be above the P-side proton electrochemical potential: $U_0/2
> \mu_{\rm P} = V_p/2,$ in order to avoid discharging the proton battery
through the shuttle. However, to allow loading the shuttle with the first electron, the energy, $U_0/2$, of the electron-binding
sites on the shuttle should be near or below the potential, $V_e/2$, of the source: $U_0/2 \leq V_e/2$. This means that the
charging energy $U_0$ of the shuttle should fit the following transport window: $V_e \geq U_0
> V_p$, in order for the pump to work.

The last cluster N2 of Complex I \cite{Schultz01,Verkh08}, having a midpoint redox potential $ E_{m7}(N2)$ = $-$150 mV , can be
considered as a source of electrons, and the Fe$_2$S$_2$-cluster of the complex III \cite{Trump81,Brandt96}, with a redox
potential $ E_{m7}({\rm Fe}_2{\rm S}_2) \simeq$ 280 mV,  is regarded as an electron drain.
The difference between the redox potentials $ E_{m7}({\rm Fe}_2{\rm S}_2)$ and $ E_{m7}(N2)$ gives a rough estimate of the
source-drain electron voltage buildup, $V_e \simeq$ 430 meV. For the charging energy we select the value  $U_0$ = 450 meV, which
slightly exceeds the electron potential difference $V_e \sim 400 $ meV, and it is significantly higher than the proton voltage
buildup $V_p \sim 250 $ meV \cite{Alberts02,Lodish07}. We find that this choice, hampering the electron transport, results in
almost equal electron and proton populations of the shuttle, and, thus, in a more efficient proton pumping across the inner
membrane.

\subsection{Pumping mechanism}
The pumping mechanism can be illustrated by considering a single electron-proton pair (see Fig.~2). In this case the shuttle has
one electron binding site (one electron potential well, shown in dark red in the upper part of each panel), and one proton site
(one proton potential well, shown in grey in the lower part of each panel). The electron and proton charges of this pair almost
completely compensate each other, and, because of this, they have practically no effect on the other electron-proton pair which
can occupy the same ubiquinone molecule. In our example each electron or proton potential well has a single energy level
(so-called ``active" level, see black continuous lines crossing the wells). The dashed black line denotes the same electron or
proton energy level, but shifted by the value of the electrostatic electron-proton energy $U_0$. This ``virtual" level can be
turned into the active level, and thus can be occupied, depending on the occupational state of the other well on the shuttle. To
illustrate this, we introduce two possible energy scales, $E_{e1},\, E_{e2},$ for an electron and two energies, $E_{p1},\,
E_{p2},$ for a proton, residing on the shuttle. The higher electron (proton) energy  $E_{e1}\; (E_{p1})$ points to the energy
level of the electron (proton) for the case when the adjacent proton (electron) well is empty: $E_{e1} = E_e = U_0/2 \ (E_{p1} =
E_p = U_0/2).$ The lower energy scale $E_{e2}\; (E_{p2})$ gives the energy of an electron (proton) when the nearby proton
(electron) potential well is occupied: $E_{e2} = E_e - U_0 = -U_0/2 \  (E_{p2} = E_p - U_0 = -U_0/2).$

The shuttle, described by the position $x$, diffuses between the points  $\pm x_0$. At $x = + x_0$ (upper left corner of loop 2 in
Fig.~1) the electron potential well is coupled to the source (S) and the proton well is coupled to the N-side of the membrane
(Fig.~2a-2c). At $x = - x_0$ (bottom right of Fig.~1) the shuttle is close to the P-side of the membrane and to the electron drain
(D) (Fig.~2d-2f). In Fig.~2 the sites populated with an electron or with a proton are shown as filled circles, whereas the white
circles show the empty sites. According to quantum transport theory \cite{Wingr,NEMS,RobPRB08}, an electron (proton) can jump to
the shuttle from the nearby electron (proton) reservoir if the electrochemical potential of the reservoir is comparable or higher
than the active energy level of the electron (proton) on the shuttle. Unloading the shuttle with an electron (proton) to the
nearby reservoir occurs in the opposite situation: when the electron (proton) energy level on the shuttle is higher than the
potential of the electron (proton) reservoir.

In the proton pumping regime, when $V_e \geq U_0 > V_p > 0$, the electrochemical potential, $\mu_{\rm S} = V_e/2$, of the electron
source is close to or higher than both possible electron energies $E_{e1}$ and $E_{e2}$ on the shuttle: $\mu_{\rm S} \geq E_{e1} >
E_{e2}$. The electron potential of the drain,  $\mu_{\rm D} = - V_e/2$, is lower than the energy of the electron, regardless of
the proton population of the shuttle: $\mu_{\rm D} \leq E_{e2} <E_{e1}.$  Therefore, the empty shuttle is always ready to accept
an electron from the source, and the shuttle, populated with an electron, is ready to donate this electron to the drain. This
means that the electron transport from the source to the drain is gated by the mechanical motion of the shuttle only, not by
protons. In contrast, since $\mu_{\rm N} = - V_p/2 <  E_{p1}$, a proton can move to the shuttle from the N-side of the membrane
only when the shuttle is populated with an electron, so that the proton energy is lower than the electrochemical potential of the
N-side: $E_{p2} < \mu_{\rm N}.$ The unloading of the proton cargo of the shuttle to the P-side of the membrane also occurs only
when the electron previously escapes from the shuttle to the drain, so that the proton energy level increases above the
electrochemical potential of the P-side: $E_{p1} > \mu_{\rm P} = V_p/2.$ At this stage the electron energy is converted to the
energy of the proton on the shuttle. Thus, besides the mechanical motion, the proton transfer from the matrix (N) to the
intermembrane space (P) is essentially gated by the electrons. This gating prevents short-circuiting the proton transfer reaction,
and, in addition, provides an efficient proton pumping across the inner mitochondrial membrane.

The pumping cycle starts (see Fig.~2a) when the shuttle is near the N-side of the membrane and near the electron source S ($x = +
x_0$). In Fig.~2b an electron jumps from the S-lead to the shuttle, since $\mu_{\rm S} \sim E_{e1}$, thus lowering the proton
energy on the shuttle from $E_{p1}$ to the level $E_{p2} < \mu_{\rm N}$. In Fig.~2c a proton moves from the N-side to the shuttle,
decreasing the electron energy to the value $E_{e2}$. Now the fully occupied neutral molecular shuttle diffuses to the opposite
side of the membrane (from Fig.~2c to Fig.~2d). In Fig.~2d the shuttle is close to the P-side of the membrane and to the electron
drain (D) ($x = - x_0$). In view of the relations $E_{e2} > \mu_{\rm D}$ and  $E_{p2} < \mu_{\rm P}$, the electron leaves the
shuttle first, escaping to the drain reservoir (see Fig.~2e). Without the Coulomb attraction, the proton energy is increased above
the level $E_{p1}$, which is higher than the electrochemical potential $\mu_{\rm P}$ of the P-side. In Fig.~2f the proton jumps to
the P-side, and after that the empty shuttle returns back to the N-side of the membrane and to the S-lead, thus completing the
pumping cycle. As a result of this cycle, one electron is transferred downhill from the source (Fig.~2a) to the drain (Fig.~2f),
and one proton is pumped uphill from the N (Fig.~2a) to the P-side (Fig.~2f) of the membrane. A similar mechanism describes the
concerted transfer of two electrons and two protons (see Fig.~3 in the next section).

\subsection{Results}

To quantitatively describe the pumping process, we numerically solve the system of master equations together with the Langevin
equation (see Eqs.~(1,2) in \textit{Methods}) for a parameter regime which provides a robust and efficient proton pump, and also
corresponds to the ubiquinone molecular shuttle, Q$_{10}$ = C$_{59}$H$_{90}$O$_4$. This molecule, which has the ability to bind
two electrons and two protons, is widely distributed in the inner mitochondrial membrane of animal cells.

In Fig.~3 we present the time evolution of the shuttling and pumping process for $V_e$ = 400 meV, $V_p$ = 250 meV, and $T$ = 298
K. The shuttle starts its motion at $x=x_0$ (Fig.~3a) and after that diffuses between the membrane borders (shown by two dashed
red lines at $x = \pm \, 2$ nm). On average, the shuttle crosses the membrane (the distance $2x_0 = $ 4 nm) in a time $\Delta t
\sim 2.7~\mu$s, which is very close to the characteristic diffusion time $\Delta t = (2x_0)^2/(2D)$ for $D = 3\cdot
10^{-12}$~m$^2$/s.

We find that, due to the symmetric configuration of the shuttle,
 both electron sites on the shuttle are populated and depopulated in concert: $n_{e1}(t) = n_{e2}(t).$
 The same applies for the proton sites, $n_{p1}(t) = n_{p2}(t).$ Fig.~3b shows the time dependence of the total electron ($n_e
= n_{e1}+n_{e2}$, continuous blue line) and proton ($n_p = n_{p1}+n_{p2}$, dashed green line) populations of the shuttle. At the
beginning of the process (at $t\sim 0$ and $x \sim x_0$) the electron sites $1_e$ and $2_e$ are occupied simultaneously, with
total population $n_e \sim 2$. As a result of the concerted reduction of the shuttle with two electrons \cite{Osyczka04}, no
intermediate semiquinone molecule, having a single unpaired electron, is formed. This conclusion is in agreement with the results
of Ref.~\cite{Verkh08}, where no semiquinone radical was detected in the reaction of the ubiquinone reduction by the tetranuclear
cluster N2 of Complex I. With a tiny delay, the proton sites  are also populated in concert.

The almost neutral ubiquinol molecule QH$_2$, loaded with approximately two electrons and two protons, diffuses and eventually
reaches the P-side of the membrane (at $t \sim 3 \, \mu$s, see Fig.~3a), where both electrons and both protons are transferred to
the low-energy drain lead or to the high-energy P-side of the membrane, respectively (Fig.~3b). It follows from Fig.~3c, that
changing the shuttle's populations in time is synchronized with the electron transfer from the source to the drain, described by
the number $N_{\rm D}$ (continuous blue line), as well as with the uphill proton translocation from the negative to the positive
side of the membrane characterized by the proton number $N_{\rm P}$ (dashed green line). The empty and neutral quinone molecule
diffuses back, to the opposite side of the membrane (Fig.~3a), and the process starts again. Notice that, as a consequence of the
stochastic nature of the Brownian motion, the shuttle is not populated and depopulated completely, and the proton population is a
little bit smaller than the electron population of the shuttle (see Fig.~3b). Because of this, the shuttle acquires a tiny charge
which can slightly increase the drag coefficient $\zeta$. Here we ignore these changes.

In Fig.~4 and~5 we show the average electron and proton particle currents, $I_{\rm D}~=~\langle N_{\rm D}\rangle /\tau_R,\ I_{\rm
P}~=~\langle N_{\rm P}\rangle/\tau_R$, as functions of the electron $(V_e)$ and proton ($V_p$) electrochemical gradients at
$T$~=~298 K, diffusion coefficient $D~=~3\cdot 10^{-12}$~m$^2$/s, and electrostatic energy $U_0~=~450$ meV. The numbers of
electrons, $\langle N_{\rm D}\rangle $, transferred from the source (Complex I) to the drain (Complex III) in the time $\tau_R$,
and the number of protons, pumped from the N- to the P-side of the membrane in the same time $\tau_R$, are numerically averaged
here over several Brownian trajectories. Positive values of the electron current, $I_{\rm D} >0$, correspond to the downhill (from
source to drain) flow of electrons, whereas positive values of the proton current, $I_{\rm P} >0$, are related to the uphill
translocation of the protons, from the N-side of the membrane (matrix) to the P-side (intermembrane space). At low values of $V_e$
and $V_p$: $V_e < 200$ meV, $V_p < 200 $ meV, both electron and proton currents are small (Figs.~4 and 5). The pumping of protons
against the potential gradient $V_p \sim 200$ meV becomes possible at high enough electron potential gradients, $V_e > 300 $ meV
(see the red area in Fig.~5), if $V_p < V_e$. At $V_e = 400$ meV the system using a single ubiquinone shuttle can pump $\sim$~200
protons in one millisecond against the transmembrane potential $V_p \sim 250 $ meV. This process is accompanied by a downhill
electron flow (the red region in Fig.~4). Both electron and proton currents saturate at $V_e > 450$ meV, if $V_p < 200$ meV. It
can be expected that the electron and proton currents across the membrane should grow almost linearly with increasing the number
of the shuttles in the ubiquinone pool.

For a sufficiently high electron driving force, $V_e >$ 450 meV, the electron current remains positive, $I_{\rm D} > 0$, even for
a high proton voltage buildup, $V_p \sim 500 $ meV, whereas protons move downhill with $I_{\rm P} < 0.$ However, the proton
component is able to \textit{reverse} the electron flow at high proton gradients, $V_p > 300 $ meV, when the electron voltage is
low enough, $V_e < 250$ meV (blue area in Fig.~4). The reversibility of the electron flow in Complexes I and III in the presence
of high proton transmembrane potentials has been observed in many respiratory and photosynthetic systems
\cite{Brandt06,Osyczka04,Armen05,Miki94}. In the reverse regime, the pump is able to translocate near 250 electrons per
millisecond against the potential gradient of $V_e \sim $ 200 meV, if the proton potential gradient is $V_p \sim 400 $ meV.
Figure~4 shows also that at sufficiently large electron driving forces ($V_e > 450$ meV) the flow of electrons cannot be reversed
($I_{\rm D} > 0$) even by the very large transmembrane proton potential ($V_p \sim 500$ meV).

 The temperature dependence of the electron, $I_{\rm D}$, and proton, $I_{\rm P}$, currents, at $V_e$ = 400 meV and $V_p$ = 250
 meV,  is presented  in Fig.~6 for $V_e = 400 $ meV, $V_p = 250 $ meV, and at three values
of the drag coefficient $\zeta$ of the shuttle: $\zeta_0/2$ (blue continuous line, low viscosity), $\zeta_0$ (green dashed line,
intermediate viscosity), and $2 \zeta_0$ (red dash-dotted line, high viscosity of the inner mitochondrial membrane). Here the drag
coefficient $\zeta_0 = 1.37$~nN$\cdot$s/m corresponds (at $T$ = 298 K) to the ubiquinone diffusion coefficient $D = T/\zeta =
3\cdot 10^{-12}$~m$^2$/s measured in Refs. \cite{Chazotte91,Marchal98}. In order to gain further insight into the physical
mechanism of the proton pump, we expand our plot to the two extreme regimes: very low and very high temperatures.
 It follows from Fig.~6a that for membranes with high viscosity
($\zeta = \zeta_0$ and $\zeta = 2 \zeta_0$) the electron transfer is accelerated and finally saturated with increasing temperature
(Fig.~6a). In a warm environment, the shuttle performs more trips from the source to the drain and back transferring downhill more
electrons. However, if the temperature is very high, and, especially, if the viscosity of the membrane is low ($\zeta =
\zeta_0/2$), electrons do not have enough time to populate or depopulate the ubiquinone molecule. Thus, in spite of the more
frequent trips, the shuttle carries less cargo, and the electron current decreases for very high temperatures and when $\zeta =
\zeta_0/2$. This phenomenon is most conspicuous in the case of proton transfer (Fig.~6b). Loading (unloading) the shuttle with
protons follows its loading (unloading) with electrons. Because of this, within the present model, the mobile shuttle moving in a
warm medium has more chances to carry less protons, thus, slowing the proton transfer at high temperatures $T \geq 400$~K. As in
the electron case, an initial increase of the proton current $I_{\rm P}$ with temperature reflects more frequent trips of the
shuttle between the membrane sides. Note that, at any viscosity of the membrane, the uphill proton current has a smooth maximum
near the physiological (or body) temperatures, $T$~=~310~K~=~37$^\circ$C.

\newpage

\subsection{Conclusions}

We have shown that the electrostatic attraction between electrons and protons, travelling together on the molecular shuttle, can
be responsible for the pumping of protons against the transmembrane electrochemical potential in the second loop of the
respiratory chain (between Complexes I and III). The Brownian shuttle (a ubiquinone/ubiquinol molecule) has two electron-binding
and two proton-binding sites. We have numerically solved a set of master equations, which quantitatively describes the process of
loading and unloading the shuttle with electrons and protons, along with a stochastic Langevin equation for the shuttle position.
We have found that, depending on the ratio between the electron potential gradient $V_e$ and the difference $V_p$ of proton
electrochemical potentials across the membrane,  the system can work \textit{either as a proton pump or, in the reverse regime, as
an electron pump} (Figs.~4 and 5). We have also studied the temperature dependence of the proton pumping for membranes with
various viscosities and found that the uphill proton current is maximal near the body temperature $T = 37 ^{\circ}$C (Fig.~6).

\subsection{Methods}

\textbf{Occupational states of the shuttle}. The occupational states of the shuttle is represented by a set of basis functions,
$|e_1 e_2\ p_1 p_2\rangle$, where the occupations of the electron and proton sites $e_1,e_2,p_1,p_2$ take the values 0 or 1. The
vacuum state, $|1\rangle = |00\ 00\rangle$, corresponds to the empty shuttle (ubiquinone), and the shuttle loaded with two
electrons and two protons (ubiquinole) is described by the state $|16\rangle = |11\ 11\rangle.$ These basis states, $|\mu\rangle
\;(\mu=1,\ldots,16)$,  and the spectrum of their eigenenergies, $\{ E_{\mu}\}$,  are closely related to the electron-proton basis
states and the energy spectrum introduced in Ref.~\cite{ASPRE08} for the proton pump in cytochrome $c$ oxidase. Here we do not
consider the spin degrees of freedom of electrons and protons.

\textbf{Master equations describing the population of the shuttle.}
 The probability of finding the system in the occupational state $|\mu\rangle $ at the time $t$ is determined by the diagonal element
 $\rho_{\mu}(t)$ of the density matrix, averaged over the Fermi distributions of the electron
 and proton  reservoirs. For these probabilities, we derive a system of master equations
\cite{RobPRB08,ASPRE08,Haken04}:
\begin{equation} \dot{\rho}_{\mu} = \sum_{\nu} \gamma_{\mu \nu}(x)\,  \rho_{\nu} - \sum_{\nu}\gamma_{\nu \mu}(x)\,  \rho_{\mu}
, \end{equation} with the relaxation matrix,
\begin{eqnarray}
\gamma_{\mu\nu}(x) = \sum_{\alpha=S}^D \sum_{\sigma=1}^2\Gamma_{\alpha\sigma}^e |w_{\alpha}(x)|^2 \{ |a_{\mu\nu}^{\sigma}|^2 [ 1 -
f_{\alpha}(\omega_{\nu\mu})] + \nonumber\\
|a_{\nu\mu}^{\sigma}|^2 f_{\alpha}(\omega_{\mu\nu}) \} + \nonumber\\
\sum_{\beta=N}^P \sum_{\sigma=1}^2\Gamma_{\beta\sigma}^p |W_{\beta}(x)|^2 \{ |b_{\mu\nu}^{\sigma}|^2 [ 1 -
F_{\beta}(\omega_{\nu\mu})] + \nonumber\\ |b_{\nu\mu}^{\sigma}|^2 F_{\beta}(\omega_{\mu\nu}) \},
\end{eqnarray}
which depends on the position $x$ of the shuttle and on the frequencies $\omega_{\mu \nu} = E_{\mu}-E_{\nu}\  (\hbar = 1)$. Here,
$ \Gamma_{\alpha\sigma}^e = \Gamma_e $ and $\Gamma_{\beta\sigma}^p = \Gamma_p$ are the energy-independent rates for electron and
proton tunneling between the binding sites on the shuttle $(\sigma =1,2)$ and the corresponding electron ($\alpha$~=~S,D) and
proton ($\beta$~=~N,P) reservoirs. Functions $w_{\alpha}(x)$ and $W_{\beta}(x)$ describe the dependence of the electron and proton
transition rates on the position of the shuttle $x$. Also, $a_{\mu\nu}^{\sigma}$ and $b_{\mu\nu}^{\sigma}$ represent matrix
elements of the electron, $a$, and proton, $b$, annihilation operators, related to the electron or proton $\sigma-$sites on the
shuttle, and $ f_{\alpha}(E), F_{\beta}(E)$ are the Fermi distribution functions in the electron, $f$, and proton, $F$,
reservoirs. The total probability is equal to one: $\sum_{\mu=1}^{16} \rho_{\mu}$~=~1. The populations of electron-binding sites,
$n_{e1}, n_{e2},$ and proton-binding sites, $n_{p1}, n_{p2},$ of the shuttle can be expressed in terms of the functions
$\rho_{\mu}$. For example, for the population of the first electron site on the shuttle we obtain: $n_{e1} =
\rho_2+\rho_6+\rho_7+\rho_{10}+\rho_{11}+\rho_{12}+\rho_{14}+\rho_{16}$.

Using tools from statistical mechanics, we model the electron source and drain reservoirs by Fermi distributions over energy $E$
with temperature $T\ (k_B=1)$ and electrochemical potentials $\mu_{\rm S}$ for the source and $\mu_{\rm D}$ for the drain:
$f_{\alpha}(E) = \{\exp[(E-\mu_{\alpha})/T]+1\}^{-1} \;(\alpha = {\rm S,D})$. The values of the chemical potentials, $\mu_{\rm S}$
and $\mu_{\rm D}$, provide the maximum energies of electrons in the last cluster of Complex I and in the first cluster of Complex
III, respectively.  Similarly, protons in the matrix (N-side) and in the intermembrane space (P-side) are described by the Fermi
distributions, $F_{\beta}(E) = \{\exp[(E-\mu_{\beta})/T]+1\}^{-1} \;
 (\beta = {\rm N,P}),$ with the corresponding proton electrochemical potentials
$\mu_{\rm N}$ and $\mu_{\rm P}$. We choose symmetric configurations for the electrochemical potentials: $\mu_{\rm S} = V_e/2,
\mu_{\rm D} = - V_e/2,$ and $\mu_{\rm N} = -V_p/2, \mu_{\rm P} = V_p/2$. The parameter $V_e$ is the difference in electrochemical
potentials between the source (Complex I) and the drain (Complex III), and $V_p$ is the proton electrochemical gradient across the
membrane. As mentioned before, the absolute value of the electron charge $|e|$ is included to the definitions of electron and
proton voltage drops $V_e$ and $V_p$, which are measured in meV. The positive values of the electron and proton voltages
correspond to the case when $\mu_{\rm S}
> \mu_{\rm D}$ and $\mu_{\rm N} < \mu_{\rm P}.$

Here we assume that the population and depopulation of the shuttle occurs via the tunneling of electrons, characterized by the
tunneling length $\lambda_e$. The Q$-$binding site of Complex I (the source of electrons) is located near the N-side of the
membrane, with coordinate $x = x_0$, and the Q$-$binding site of Complex III (the electron drain) is close to the P-side, with $x
= - x_0$. In this case we use the standard (in nanoelectromechanics, \cite{NEMS}) expressions describing the position-dependence
of the source-shuttle and drain-shuttle electron-tunneling amplitudes: $w_{\rm S}(x)~=~\exp(-\,|x-x_0|/\lambda_e), \ w_{\rm
D}(x)~=~\exp(-\,|x+x_0|/\lambda_e)$. The region $x > x_0$ corresponds
 to the matrix (N-side), and the region with $x < - x_0$ refers to the intermembrane space (P-side).
Protons from both, N and P, aqueous sides of the inner membrane can move along proton-conducting pathways \cite{Hunte03} and reach
the Q-binding sites of Complexes I and III.  Consequently, a proton exchange between the shuttle and the aqueous intermembrane
space is possible when $x > x_0$, and the proton transitions between the P-side and the shuttle take place when $x < - x_0$. This
situation is modelled here with the position-dependent proton amplitudes: $W_{\rm N}(x)~=~[ \exp((x_0-x)/\lambda_p)+1 ]^{-1}$, and
$W_{\rm P}(x)~=~[ \exp((x+x_0)/\lambda_p)+1 ]^{-1}$, characterized by the steepness $\lambda_p$. These forms quantify the above
observations.

The transition rate $\Gamma_e$, between the electron-binding sites and the electron reservoirs, is assumed to be about the proton
transition rate $\Gamma_p $ (between the aqueous sides of the membrane and the proton-binding sites on the shuttle): $\Gamma_{e}
\simeq \Gamma_{p} \simeq 5\ $ns$^{-1}$. For these rates the shuttle can be loaded (or unloaded) with electrons and protons in a
time which is much shorter than the diffusion time scale, $\Delta t \simeq (2 x_0)^2/(2D)~\sim 2.7 \, \mu$s, of the system. Here
the distance $2 x_0 = 4 $ nm is close to the membrane thickness. The electron tunneling length $\lambda_e$ and the proton
steepness parameter $\lambda_p$ are chosen to be $\lambda_e = \lambda_p =0.25$ nm. Protons can be delivered to the Q$-$docking
sites by a system of proton-conducting pathways \cite{Hunte03}. We assume that the shuttle is loaded with N-side protons at $x >
(x_0-\lambda_p)$, and unloaded in the  P-side protons at $x<(-x_0+\lambda_p)$, where $\lambda_p = $ 0.25 nm. The Q$-$docking sites
are located on the opposite sides of the inner membrane and are separated by the distance $2x_0$.

\textbf{Stochastic equation describing the shuttle motion}. Electron and proton transitions between the shuttle and the
corresponding reservoirs have negligible effect on the mechanical motion. However, as follows from Eq.~(2), electron and proton
transition rates drastically depend on the spatial gap between the shuttle and the leads. The one-dimensional diffusion of the
Brownian shuttle along a line, connecting the quinone-binding site of Complex I and the quinone-binding site of Complex III, is
described by the overdamped Langevin equation:
\begin{equation} \zeta \dot{x} = -\, \frac{dU(x)}{dx} + \xi,
\end{equation}
where $\zeta$ is the drag coefficient of the shuttle in the lipid membrane. A zero-mean valued, $\langle \xi\rangle = 0,$
fluctuation force $\xi$ has Gaussian statistics with the correlation function: $\langle \xi(t)\xi(t')\rangle = 2 \zeta T \delta
(t-t'),$ which is proportional to the temperature $T$ of the environment. A potential $U(x)$ is responsible for the spatial
confinement of the ubiquinone/ubiquinole molecule inside the inner mitochondrial membrane. The diffusion coefficient of
ubiquinone, $D~=~T/\zeta \sim 3\cdot 10^{-12}$~m$^2$/s, measured in \cite{Chazotte91,Marchal98} at $T$~=~298 K , corresponds to
the drag coefficient $\zeta_0~=~1.37$~nN$\cdot$s/m. We use the value of $\zeta_0$ as a reference point (see Fig.~6).

\subsection{Acknowledgements}
This work was supported in part by  the National Security Agency (NSA), Laboratory of Physical Science (LPS), Army Research Office
(ARO), National Science Foundation (NSF) grant No. EIA-0130383, JSPS-RFBR 06-02-91200, and Core-to-Core (CTC)
 program supported by the Japan Society for Promotion of Science (JSPS).
 S.S. acknowledges support from the EPSRC via EP/D072581/1.

\begin{figure}
\includegraphics[width=16.0cm, height=22.0cm ] {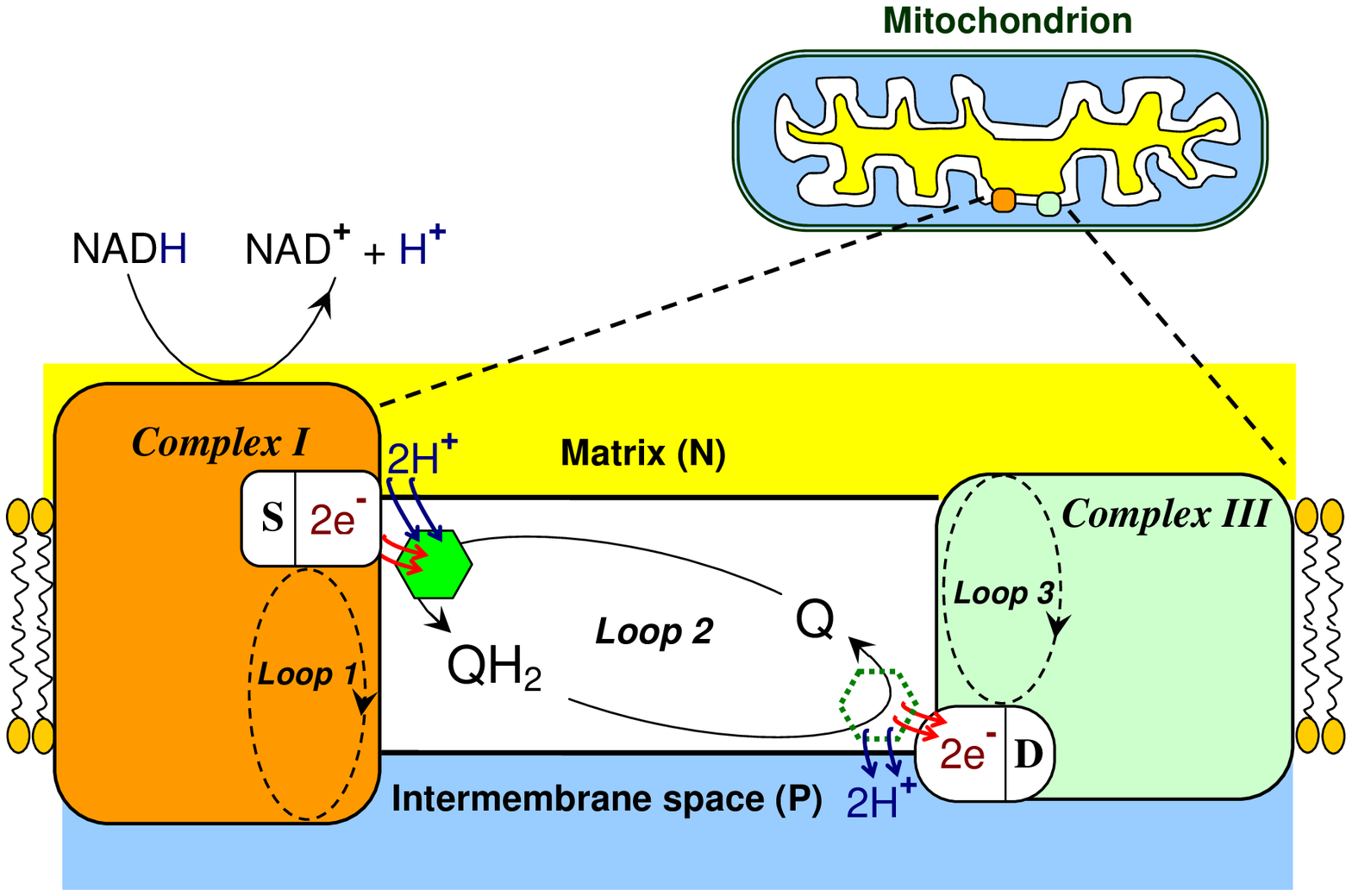}
\vspace*{-8cm} \caption{ (Color online) \textbf{Schematic diagram of the second loop of the respiratory chain.} High-energy
electrons are delivered to Complex I by NADH and forwarded to the ubiquinone molecule (Q) shuttling between Complexes I and III.
Complex I serves as a source (S) of electrons, and Complex III plays the role of an electron drain (D). In addition to two
electrons, the shuttle carries two protons taken from the matrix side (N) of the inner membrane at $x=+x_0$. This fully loaded
shuttle (the ubiquinol molecule QH$_2$) diffuses freely to the P-side of the membrane, at $x=-x_0$, and donates two electrons to
Complex III. Thus, in this process, two protons translocate energetically uphill to the intermembrane space (P).}
\end{figure}

\begin{figure}
\includegraphics[width=12.0cm]{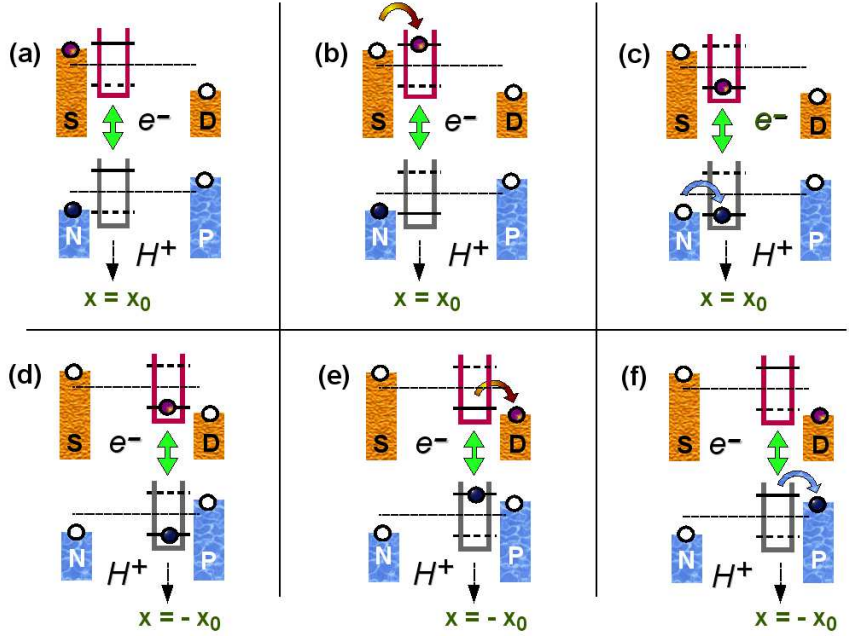}
\vspace*{-1cm} \caption{ (Color online) \textbf{Energetic diagrams of the pumping process}. The shuttle schematically shown here
carries a single electron and a single proton: S and D are the Source and Drain reservoirs for electrons (rust-orange), N and P
are the Negative and Positive aqueous sides (blue) of the inner membrane. Of course, in reality the shuttle carries two electrons
and two protons. However, it is simpler to explain the proton pumping process with only one of each. Potential wells for electrons
(dark red, upper part of each panel) and for protons (grey, lower part) move with the shuttle from one side of the membrane
($x=+x_0$) to the other side ($x=-x_0$). In (a,b) the shuttle is near the N-side, loading an electron from the source (S). In (a)
the energy level of the proton [see the lower part of (a)] is higher than the electrochemical potential of the N-side. In (b) the
shuttle is populated with one electron. The effective energy of the proton is now lower than the potential of the N-side, and the
shuttle is then loaded with the proton in (c). The fully loaded shuttle in (c) is still located near the N-side of the membrane
($x=+x_0$). Both, electron and proton, lower their energies because of the electron-proton Coulomb attraction (vertical green
arrows). (d) The diffusing shuttle carries its cargo to the P-side of the membrane at $x=-x_0$. The energy level of the electron
on the shuttle is now higher than the electron potential of the drain, and the electron translocates to the drain in (e). Without
the electron-proton Coulomb attraction the energy level of the proton on the shuttle goes up, over the electrochemical potential
of the P-side, and the shuttle gives up its proton in (f). Thus, the proton is pumped to the P-side of the membrane, in the lower
part of (f), and the electron is in the drain. The shuttle is now empty and ready to diffuse back, to the N-side of the inner
membrane, in order to start the pumping cycle again. }
\end{figure}

\begin{figure}
\includegraphics[width=26.0cm, height=15cm ]{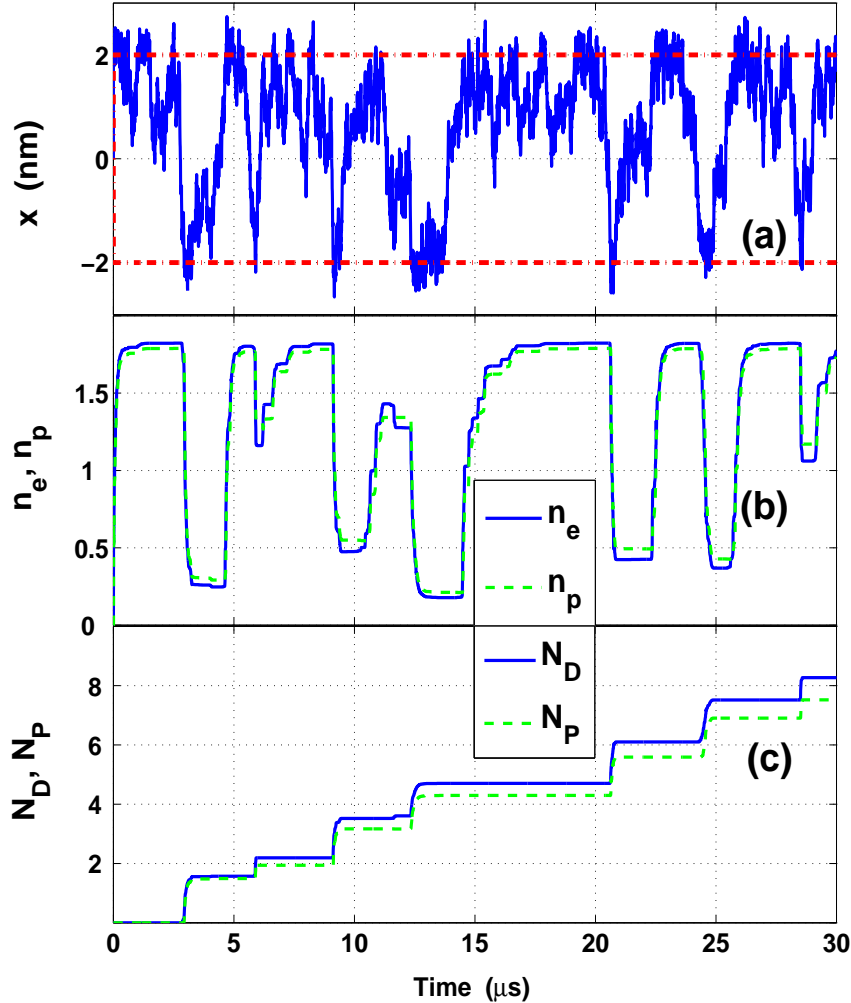}
\vspace*{-1cm} \caption{ (Color online) \textbf{Time evolution of the pumping process}. (a) Time dependence of the position $x$
(in nm) (blue continuous curve) of the ubiquinone (ubiquinol) molecule Q (QH$_2$), shuttling between the walls of the inner
mitochondrial membrane located at $x = \pm x_0$ (red dashed lines), where $x_0$ = 2 nm;
 (b) the total electron ($n_e = n_{e1} + n_{e2}$, blue continuous line) and proton ($n_p = n_{p1} +
n_{p2}$, green dashed line) populations of the shuttle versus time (in $\mu$s);
 (c) the number of transferred electrons ($N_{\rm D}$, blue continuous line) and the number of pumped protons ($N_{\rm P}$, green dashed line)
 versus time at $V_e$ = 400\ meV,\ $V_p$ = 250\ meV, $\zeta = \zeta_0,$ and at $T$ = 298 K. Notice that the shuttle is loaded near the N-side
 of the membrane, at $x \approx +~x_0$, and unloaded at the P-side, at $x \approx -~x_0$. It follows from (c) that the process of shuttle
 unloading is accompanied  by a stepwise increase of the number of electrons, $N_{\rm D}$, transferred to the drain,
 and the number of protons,   $N_{\rm P}$, pumped to the P-side of the membrane. }
\end{figure}

\begin{figure}
\includegraphics[width=22.0cm ]{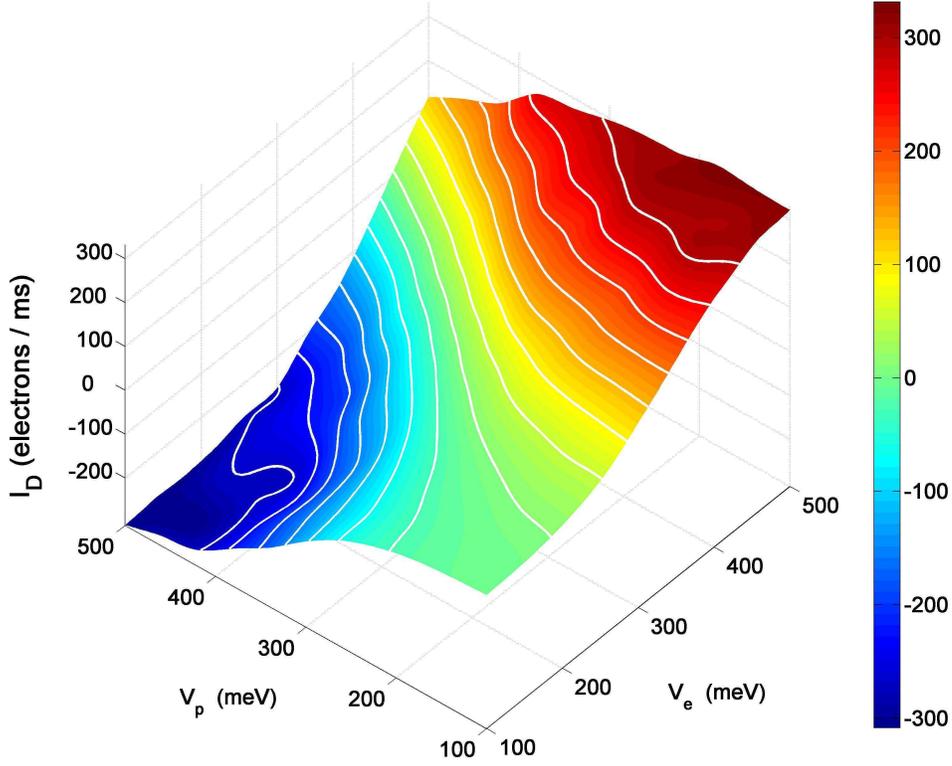}
\vspace*{1cm} \caption{ (Color online) \textbf{Electron current-voltage characteristics.} The \textit{electron} current $I_{\rm
D}$ (number of electrons transferred from the source to the drain per one millisecond) as a function of electron and proton
electrochemical gradients, $V_e$ and $V_p$, at $U_0 = 450$ meV, $\zeta = \zeta_0$, and at $T = 298 $ K. This current is positive,
$I_{\rm D} > 0$, at high enough electron voltage, $V_e
> 300$ meV, if the proton voltage is sufficiently small, $V_p < 250$ meV. In the opposite case, when $V_p > 300$ meV and $V_e < 250$
meV, the pump works in the \textit{reverse} regime, where the energy of the downhill-flowing protons (blue region in Fig.~5) is
used for the uphill translocation of electrons against the potential $V_e$ (blue area in Fig.~4).}
\end{figure}

\begin{figure}
\includegraphics[width=22.0cm ]{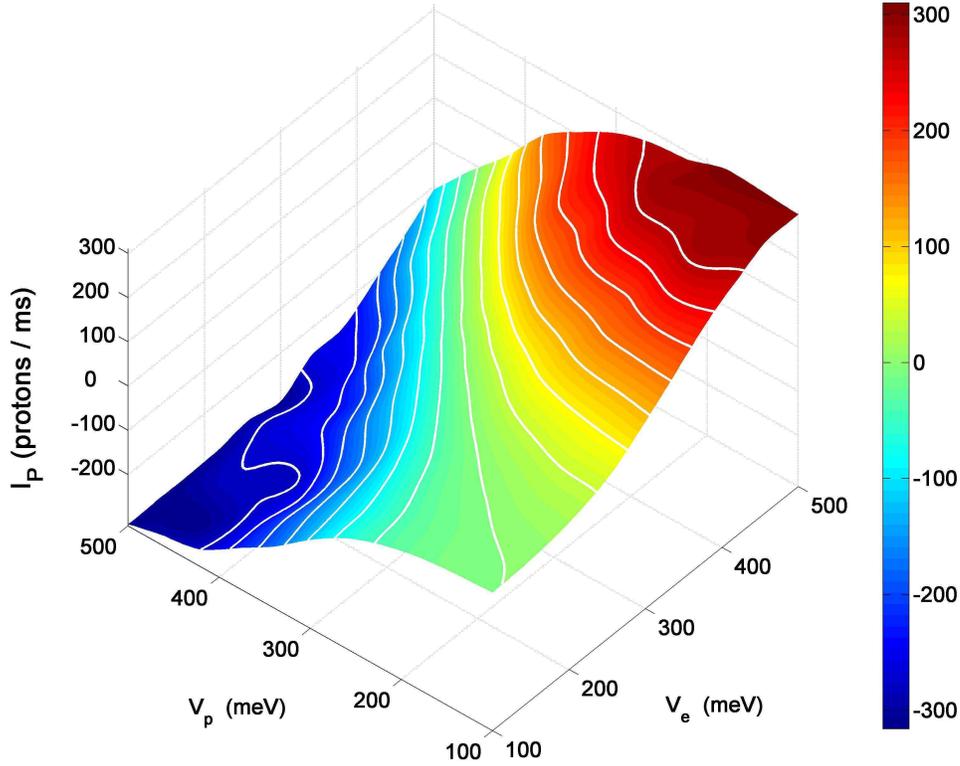}
\vspace*{1cm} \caption{ (Color online) \textbf{Proton current-voltage characteristics.} The \textit{proton} current $I_{\rm P}$
(number of protons translocated from the negative, N, to the positive, P, side of the inner membrane per one millisecond) versus
the electron, $V_e$, and proton, $V_p$, potential gradients, at $U_0 = 450$ meV, $\zeta = \zeta_0$, and $T = 298 $ K. The positive
values of $I_{\rm P}$ (red area where $V_e > 300$ meV, $V_p < 250$ meV) correspond to the regime where the protons are pumped.
Blue regions ($V_p
> 300$ meV, $V_e < 250$ meV) in Figs.~4 and 5 correspond to the electron pump regime, where protons move downhill ($I_{\rm P} < 0$),
thereby translocating electrons uphill, from the drain to the source.}
\end{figure}

\begin{figure}
\includegraphics[width=20.0cm, height=12cm ]{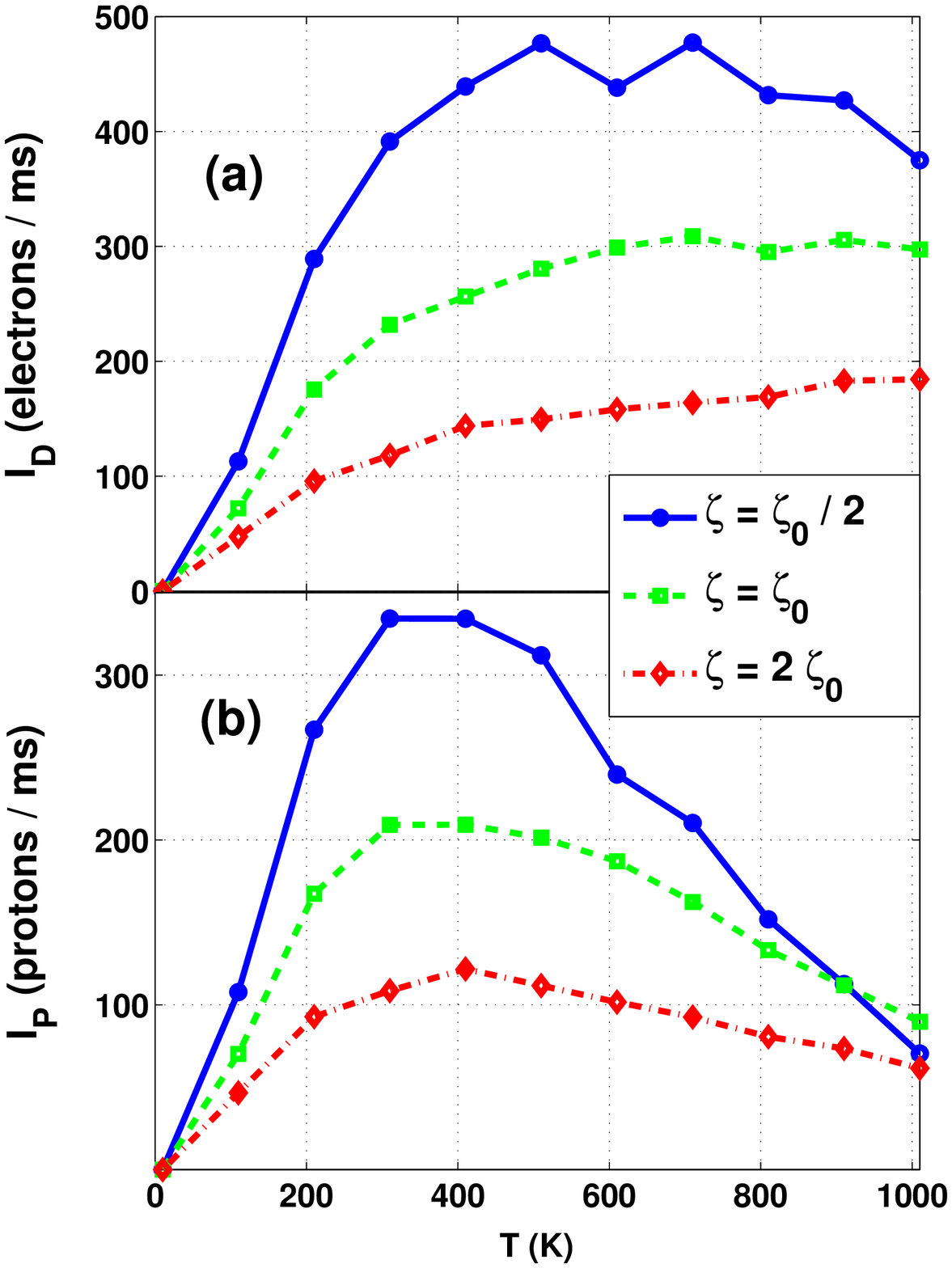}
\vspace*{-1cm} \caption{ (Color online) \textbf{Effects of temperature and viscosity}. Temperature dependence of (a) the electron
current, $I_{\rm D}$, and (b) the proton current, $I_{\rm P}$, in the proton pumping regime ($V_e = 400$ meV, $V_p = 250$ meV) for
\textit{three} levels of membrane viscosity, determined by the drag coefficient $\zeta$: (i) $\zeta = \zeta_0/2$ (low viscosity);
(ii) $\zeta = \zeta_0$ (intermediate viscosity); (iii) $\zeta = 2 \zeta_0$ (relatively high viscosity). The drag coefficient
$\zeta_0$~=~1.37~nN$\cdot$s/m is related (at $T$ = 298 K) to the diffusion coefficient $D = T/\zeta = 3\cdot 10^{-12}$~m$^2$/s
\cite{Chazotte91,Marchal98}. In the region $T \leq 250$ K, both electron and proton currents increase almost linearly with
temperature, because the shuttle performs more trips and translocates more particles as the temperature increases. At high
temperatures, $T \geq 600$ K, the electron current (Fig.~5a) saturates and even slightly decreases (at low viscosity). The proton
current (Fig.~5b) shows an even more pronounced maximum near $T \sim$ 310 K ($\sim$ 37$^{\circ}$C, body temperature), for all
values of the membrane viscosity. Protons are loaded and unloaded on the shuttle \textit{after} the electrons are loaded and
unloaded. Because of this, at high $T$ and low $\zeta$, the mobile shuttle has not enough time to be loaded or unloaded with
protons, carrying less protons across the membrane. }

\end{figure}

\end{document}